\pdfoutput=1
\documentclass[10pt, conference, letterpaper]{IEEEtran}

\usepackage{cite}
\usepackage{amsmath,amssymb,amsfonts}
\usepackage{graphicx}
\usepackage{textcomp}
\usepackage{xcolor}
\usepackage{multirow}
\usepackage{booktabs}
\usepackage{siunitx}

\providecommand{\citet}[1]{\cite{#1}}
\providecommand{\citep}[1]{\cite{#1}}

\def\BibTeX{{\rm B\kern-.05em{\sc i\kern-.025em b}\kern-.08em
    T\kern-.1667em\lower.7ex\hbox{E}\kern-.125emX}}

\begin{document}

\title{RF-Based Macronutrient Estimation in Liquid Foods via Composition-Aware Multi-Task Learning}

\author{\IEEEauthorblockN{Ishtiaque Ahmed Showmik and Kannan Athreya}
\IEEEauthorblockA{Department of Computer Science and Engineering\\
The Ohio State University, Columbus, OH, USA}
}

\maketitle

\begin{abstract}

Current approaches to macronutrient estimation range from high-overhead laboratory analysis to 
user-dependent methods such as manual nutritional lookup, reading package labels, and vision-based 
image analysis. These approaches are often time-consuming, input-dependent, or sensitive to imaging
and measurement conditions. In this work, we investigate synchronous 2.4~GHz OFDM
radio-frequency (RF) measurements for estimating carbohydrate, fat, and protein content in liquid foods.
Using a controlled sensing setup, we study how liquid composition shapes frequency-domain
RF responses---a first step toward a sensor integrated into a liquid-contacting vessel that
estimates macronutrient content passively and label-free, and toward composition-aware
RF food analysis beyond liquids. We propose the Ingredient-Conditioned OFDM Transformer (\textit{ICOT}),
a structure-aware multi-task Transformer that operates
on two-dimensional OFDM channel frequency response matrices. ICOT infers canonical ingredient
categories representing dielectric-driver composition and uses them to condition
macronutrient regression through feature concatenation. On 28{,}420 captures spanning 55 beverage
classes, ICOT achieves an in-distribution MAE of $0.101$~g/100~ml and an
out-of-distribution MAE of $0.485$~g/100~ml---comparable in magnitude to deviations
present in manufacturer-reported labels~\cite{jumpertz2013}---though generalization to
novel formulations remains the primary open challenge.
\end{abstract}


\section{Introduction}
For a person managing type~1 diabetes, miscounting the carbohydrates in a glass of juice by a few grams can precipitate an insulin dosing error with clinically significant glycemic consequences~\cite{roversi2022}.
More generally, determining what is in an unlabeled liquid requires reading a package, running a lab test, or already knowing the composition---compounded by evidence that packaged labels themselves deviate measurably from stated values~\cite{jumpertz2013}, and home-prepared beverages carry no label at all.
\citet{kirkpatrick2021} demonstrate that tracking macronutrient intake from first principles, ingredient by ingredient, imposes a compliance burden incompatible with consistent daily use.
Motivated by this gap, the long-term objective of this work is a sensor integrated into a liquid-contacting vessel that estimates the macronutrient content of any poured liquid without user involvement or prior knowledge of its composition.
In this paper, we address this objective by investigating whether active OFDM RF measurements can support composition-aware macronutrient estimation in liquid foods.

Nutrient estimation refers to quantifying carbohydrate, fat, and protein content, since these macronutrients largely determine caloric content and guide dietary decisions; existing automated approaches remain limited in several respects.
Smartphone applications~\cite{app1, app2, app4} depend on the user initiating a logging event and are sensitive to training-data coverage.
Vision-based methods~\cite{meyers2015im2calories, tanno2016deepfoodcam, xu2016development, chen2016deep} match query images against reference databases and are sensitive to viewpoint, lighting, and occlusion; most account only for food category and volume, omitting ingredient composition and cooking method.
Prior RF-based food sensing uses microwave oven leakage: WiNE~\cite{wine2022} estimates nutrient percentages during heating, and RFTemp~\cite{rftemp2021} monitors food temperature from the same signal; both require active microwave excitation and report coarse nutrient fractions rather than absolute per-volume macronutrient quantities.

As the RF signal propagates within the liquid, it is subject to attenuation and
scattering whose signature can be discerned from the spectral response---fat
droplets, dissolved sugar, and ions each manifest uniquely in the received
spectrum. In our experimental setup, we sample OFDM Channel Frequency Responses (CFR) over various channels and subcarriers using a 2.4~GHz software-defined radio with synchronized transmit and receive chains, distinct from commodity WiFi hardware.
Unlike \textit{LiquID}~\cite{10.1145/3210240.3210345}, which focuses on the nature of the liquid based on complex permittivity,
we aim to estimate concentrations of carbohydrates, fats, and proteins.
This poses a challenge: RF signals detect only the overall dielectric properties
caused by the base solvent, electrolyte ions, lipid droplets, and
non-electrolyte solutes~\cite{stogryn1971, nelson2005}. Distinct liquids
can share similar attenuation characteristics, making it impossible to
predict nutrient concentrations without first understanding what causes the
dielectric behavior.
To address this, we introduce the Ingredient-Conditioned OFDM Transformer (ICOT),
a multi-task network built on the Transformer encoder~\cite{vaswani2017} that
takes the entire 2D CFR matrix ($N_c \times N_s$ channels by subcarriers) as
input. ICOT first learns the latent space representing the main dielectric
contributors---base solvent class, molecular heterogeneity, and ion/non-ion
solutes---then infers macronutrient concentrations from this intermediate
representation; ingredient labels supervise training only within this latent
space, while inference uses RF measurements alone.

\begin{figure}[t]
    \centering
    \includegraphics[width=\columnwidth]{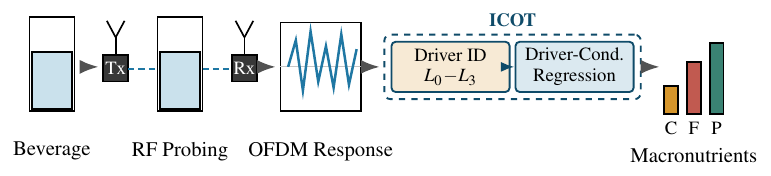}
    \caption{RF-based macronutrient estimation pipeline: an unlabeled beverage is probed with synchronized 2.4\,GHz OFDM; ICOT identifies dielectric drivers ($L_0$--$L_3$) before conditioning macronutrient regression.}
    \label{fig:overview}
\end{figure}

The primary contributions of this paper are as follows:

\begin{enumerate}
    \item We formulate macronutrient estimation in liquid foods as an indirect RF inference problem mediated by dominant dielectric drivers, and design ICOT, a multi-task Transformer decomposing this into hierarchical composition inference followed by composition-conditioned regression (canonical supervision used only in training). Unlike prior RF work's black-box sensing networks, ICOT predicts absolute macronutrient quantities in \textit{unseen} beverages by inferring latent physical structure rather than recalling prelearned identities---the basis for the interpretability analysis in Contributions 2--3.

    \item We demonstrate via embedding-shuffle intervention experiments that ingredient-level conditioning is functionally necessary---not merely additive---for in-distribution performance, and that concatenation-based fusion achieves a better accuracy--robustness tradeoff than FiLM.

    \item We evaluate ICOT on 28,420 captures across 55 beverage classes, achieving ID MAE of $0.101$~g/100ml and OOD MAE of $0.485$~g/100ml, and characterize two error sources---compositional singularity and spectral confusability---that define the path to deployment. The current system is restricted to fixed container geometry, synchronized Tx/Rx hardware, and multi-liter sample volumes; robust OOD performance remains open for compositionally novel or high-sugar formulations.
\end{enumerate}

\section{Background}\label{sec:background}
This section establishes how liquid composition governs OFDM channel responses and the implications for model design.

\subsection{Electromagnetic Propagation Through Liquid Media}
\label{sec:bg_em}
The interaction of electromagnetic waves with liquid materials is governed by 
the complex permittivity
\begin{equation}
\varepsilon^\ast(\omega) = \varepsilon'(\omega) - j\varepsilon''(\omega),
\end{equation}
where $\varepsilon'(\omega)$ controls wave velocity and phase propagation, and 
$\varepsilon''(\omega)$ captures dielectric loss and energy absorption
\cite{vonHippel1954dielectrics, nelson2003dielectric}. For a homogeneous liquid
layer of thickness $d$, the frequency-domain channel response is
\begin{equation}
H(\omega) = H_0(\omega)\,
\exp\!\left[-\left(\alpha(\omega) + j\,\beta(\omega)\right)d\right],
\end{equation}
where $\alpha(\omega)$ and $\beta(\omega)$ are the attenuation and phase
constants determined by $\varepsilon^\ast(\omega)$ \cite{rappaport1996wireless}.
Changes in liquid composition alter
$\varepsilon'(\omega)$ via solvent binding, emulsification, and molecular
polarization, producing frequency-dependent phase dispersion across
subcarriers~\cite{tse2005fundamentals}; consequently, phase
effects remain implicitly encoded in non-local spectral structure even when
magnitude-only features are used after calibration. This dual dependence on
$\varepsilon'(\omega)$ and $\varepsilon''(\omega)$ underlies OFDM's sensitivity
to liquid composition.

\subsection{Overview of OFDM}
Orthogonal Frequency Division Multiplexing (OFDM) partitions a wideband channel
into a set of orthogonal subcarriers, each probing the medium at a distinct
frequency \cite{nee2000, tse2005fundamentals}. The resulting channel response
is compactly described as discrete samples of $H(\omega)$:
\begin{equation}
\left\{ H(\omega_{c,s}) \mid c = 1,\dots,C,\; s = 1,\dots,S \right\},
\end{equation}
where $c$ is the channel index and $s$ is the subcarrier index.
This representation has two properties relevant to liquid sensing:
(i)~frequency selectivity, preserving composition-dependent dispersion and
attenuation across the band; and (ii)~inter-channel continuity from partially
overlapping adjacent channels, extending effective spectral
coverage~\cite{tse2005fundamentals}---making the measurement a structured 2D
object (channels $\times$ subcarriers) whose informative patterns may be
\emph{non-local} across frequency.

Following IEEE 802.11a/g/n/ac, a nominal $20$~MHz channel discretizes into
$N_\text{FFT}=64$ subcarriers ($N_{ST}=52$ active, DC nulled). Channels are
center-spaced at 5~MHz with 20~MHz bandwidth, so the eleven 2.4~GHz ISM-band
channels overlap and jointly cover more spectrum than any one
alone---fully supported by the WARP~v3 hardware~\cite{warpProject}.

\subsection{RF-Based Liquid Sensing}
\label{est}

\begin{figure}[t]
    \centering
    \includegraphics[width=\columnwidth]{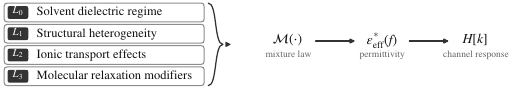}
    \caption{A liquid mixture decomposes into dielectric drivers, mapped through a
    nonlinear mixture law $\mathcal{M}(\cdot)$ to the effective complex permittivity $\varepsilon_{\mathrm{eff}}^{*}(f)$,
    governing the OFDM channel response $H[k]$.}
    \label{fig:die_drivers}
\end{figure}

Two properties motivate the learning-based framework in
Section~\ref{sec:model_overview}---(i)~frequency-dependent dielectric interactions
induce non-local spectral correlations across subcarriers, and (ii)~the
RF-to-macronutrient relationship is indirect, mediated by ingredient composition
(Figure~\ref{fig:die_drivers})---for which this section presents empirical evidence.

\begin{figure}[t]
    \centering
    \includegraphics[width=\columnwidth]{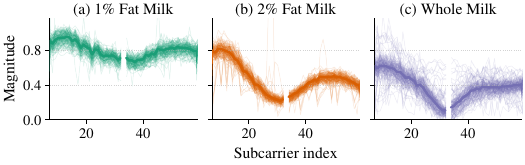}
    \caption{Baseband OFDM magnitudes for 1\%, 2\%, and Whole Milk, channel~1 (faint: individual captures; bold: per-class median; DC subcarrier omitted).}
    \label{fig:milk_channel1_subcarriers_paper}
\end{figure}

\begin{figure}[t]
    \centering
    \includegraphics[width=\columnwidth]{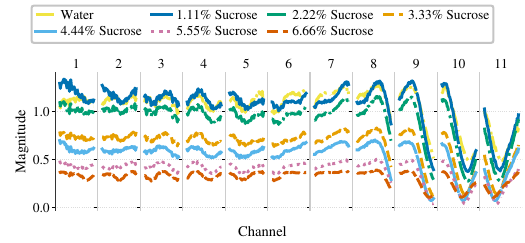}
    \caption{Calibrated OFDM magnitudes for increasing sucrose concentration; each panel is one 20\,MHz channel (64 subcarriers) of the 2.4\,GHz ISM band.}
    \label{fig:cal_agg}
\end{figure}

Figure~\ref{fig:milk_channel1_subcarriers_paper} depicts baseband OFDM magnitudes for 1\% Fat Milk, 2\% Fat Milk, and Whole Milk on channel~1: despite identical water content and setup, varying fat content noticeably shifts the magnitude spectrum, since it changes the medium's dielectric loss and produces frequency-selective attenuation $\exp(-\alpha(\omega)d)$ rather than simple scaling.

Figure~\ref{fig:cal_agg} presents air-normalized data $H_{\mathrm{norm}}(f)$ (Section~\ref{sec:eval_setup}) for sugar solutions: higher sucrose concentration yields consistently smaller normalized magnitudes and stronger spectral fluctuations, consistent with increased dielectric loss from progressive disruption of water's orientational polarization.

\begin{figure*}[htbp]
    \centering
  
        \includegraphics[width=\textwidth]{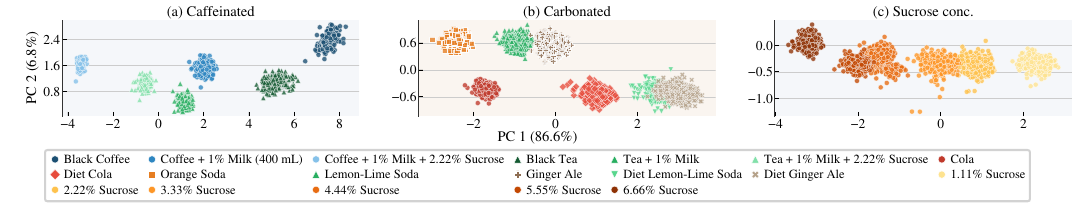}

    \caption{PCA projections of uncalibrated OFDM feature representations for
    different liquid food groups.}
    \label{fig:pca}
\end{figure*}

Figure~\ref{fig:pca} shows the resulting manifold of OFDM features: sugar water varies monotonically along the dominant principal components with dielectric loss; carbonated beverages separate by sucrose vs.\ synthetic sweeteners; and caffeinated beverages cluster by the dielectric relaxation of milk's lipid/protein content.

Rather than treating beverages as collections of discrete culinary ingredients, we model each as a liquid mixture whose RF response is governed by a small set of dominant \emph{dielectric drivers}: the effective complex permittivity $\varepsilon_{\mathrm{eff}}^{*}(f)$ at frequency $f$ is a nonlinear mixture mapping
\begin{equation}
\varepsilon_{\mathrm{eff}}^{*}(f)
=
\mathcal{M}\!\left(\left\{\varepsilon_i^{*}(f),\, c_i\right\}_{i=1}^{K},\, \mathbf{s}\right),
\end{equation}
where $\varepsilon_i^{*}(f)$ and $c_i$ are the intrinsic permittivity and concentration of constituent $i$, $\mathbf{s}$ represents state variables (temperature, mixing uniformity, microstructure), and $\mathcal{M}(\cdot)$ captures nonlinear mixing effects from heterogeneous phases, solute--solvent interactions, and dispersion phenomena. The drivers themselves are not mutually exclusive and may act simultaneously within a given mixture; their relative influence depends on both concentration and operating frequency. Critically, distinct classes of driver can be difficult to disambiguate from a magnitude-only response, since ionic conduction and non-ionic dipolar relaxation both manifest as increased attenuation despite being physically distinct loss mechanisms. This matters because the driver classes do not carry macronutrient content equally: the ionic driver ($L_2$) is typically non-caloric (e.g., electrolytes), whereas the emulsion/heterogeneity driver ($L_1$) and non-ionic solute driver ($L_3$) carry fat/protein and carbohydrate content, respectively. A direct RF-to-nutrient regressor therefore cannot attribute an observed attenuation to a nutrient-bearing versus non-nutrient-bearing driver without first resolving which class of driver is present. Section~\ref{sec:canonical_mapping} makes this decomposition operational as the four canonical levels $L_0$--$L_3$ that ICOT predicts.

\section{Proposed Model Architecture}
\label{sec:model_overview}

\begin{figure*}[t]
    \centering
    \includegraphics[width=\textwidth]{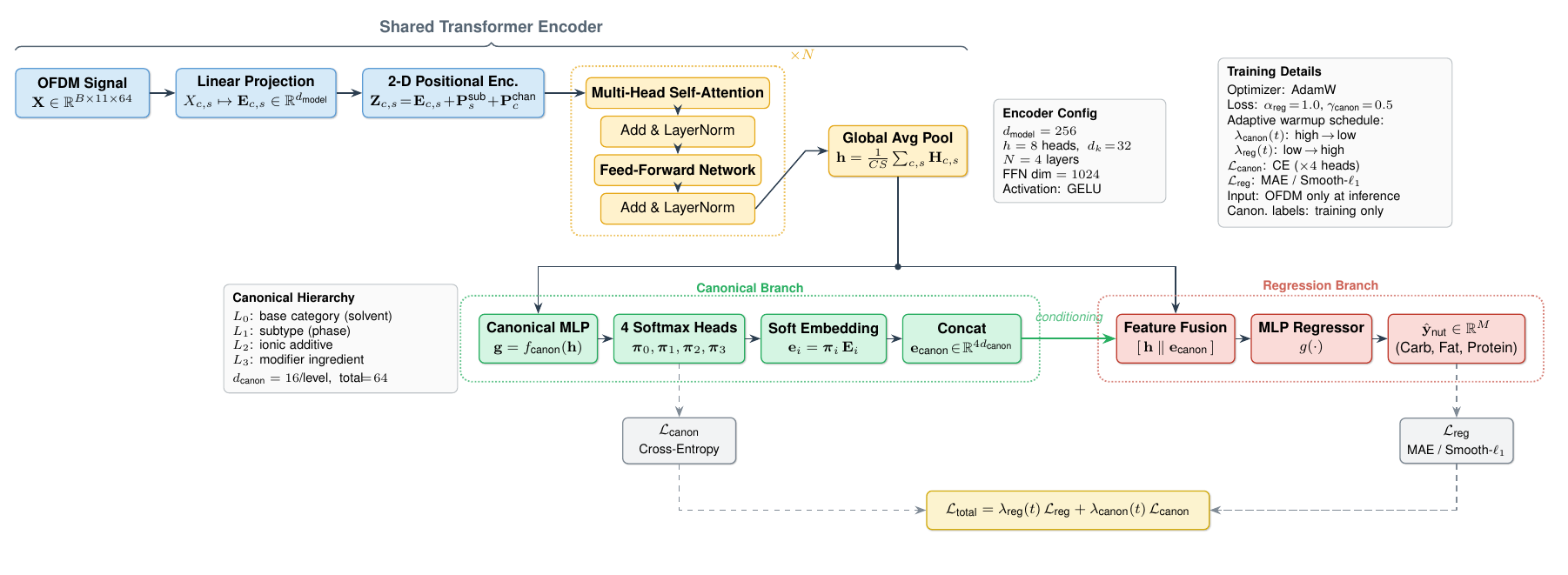}
    \caption{\textbf{Ingredient-Conditioned OFDM Transformer (ICOT)}. A shared Transformer encoder extracts a global RF feature; a canonical-structure head predicts $(L_0,L_1,L_2,L_3)$ from RF alone, and the resulting soft embeddings condition macronutrient regression. Loss weights shown are terminal values after the warmup schedule of Section~\ref{sec:joint_objective}.}
    \label{fig:icot_model}
\end{figure*}
OFDM signals transmitted through beverages carry structured, non-local
dependencies across subcarriers and channels reflecting
frequency-dependent dielectric behaviour~\cite{kaatze1989,Gregory2006}.
Since this response is composition-dependent, identical attenuation
patterns can arise from different underlying compositions, making direct
RF-to-nutrient regression ambiguous. We propose the \textit{Ingredient-Conditioned
OFDM Transformer (ICOT)}, which decomposes nutrient estimation into
(i)~\emph{canonical composition inference} and
(ii)~\emph{ingredient-conditioned nutrient regression}.

\subsection{OFDM Signal Representation}
\label{sec:signal_representation}

The input to ICOT is a calibrated OFDM channel-response tensor
\[
\mathbf{X} \in \mathbb{R}^{B \times C \times S},
\]
where $B$ is the batch size, $C=11$ the number of OFDM channels, and
$S=64$ the number of subcarriers per channel. Each entry stores the calibrated magnitude of the complex frequency-domain
response, encoding signal propagation through the liquid~\cite{Gregory2006};
phase estimates are sensitive to hardware clock offsets, so magnitude-only
features are used for robustness. The subcarrier axis captures
frequency-dependent attenuation and dielectric dispersion~\cite{kaatze1989},
while the channel axis spans partially overlapping bands with complementary
spectral views---preserving both is critical for modelling frequency-selective
and cross-channel interactions. Raw responses are normalised against a
reference air measurement to remove environmental and container-dependent
artefacts.

\subsection{Feature Embedding and Positional Encoding}
\label{sec:embedding_positional}

Each scalar subcarrier magnitude $X_{c,s}$ is projected into a
$d_{\text{model}}$-dimensional embedding $\mathbf{E}_{c,s}$ via a learnable linear
map. Because Transformers are permutation-invariant~\cite{vaswani2017},
we inject learnable positional encodings along both axes:
\[
\mathbf{Z}_{c,s} = \mathbf{E}_{c,s}
  + \mathbf{P}^{\text{sub}}_{s}
  + \mathbf{P}^{\text{chan}}_{c},
\]
where $\mathbf{P}^{\text{sub}}_{s}$ and $\mathbf{P}^{\text{chan}}_{c}$
encode subcarrier ordering and inter-channel relationships, preserving
frequency locality and spectral continuity so the encoder models
structured interactions rather than an unordered feature set.

\subsection{Transformer-Based Signal Encoder}
\label{sec:transformer_encoder}

The embedded representation
$\mathbf{Z}\in\mathbb{R}^{C\times S\times d_{\text{model}}}$ is
flattened into a sequence of length $C\!\times\!S$ and passed through a
stack of Transformer encoder layers~\cite{vaswani2017}, following the
patch-based tokenisation strategy of the Vision Transformer~\cite{Dosovitskiy2021}.
Self-attention lets every subcarrier attend to all others, capturing
long-range spectral correlations from the liquid's dielectric
properties~\cite{kaatze1989,Gregory2006}, realising a hard-parameter-sharing
multi-task scheme~\cite{Caruana1997} where the common representation is
optimised jointly for both objectives. The encoder output is aggregated
into a global RF feature $\mathbf{h}$ by average pooling over the
$C\!\times\!S$ tokens.

\subsection{Canonical Mapping Prediction and Supervision}
\label{sec:canonical_mapping}

Because the relationship between RF measurements and macronutrients is mediated by ingredient structure, we introduce a hierarchical compositional representation as an intermediate supervision target.
We represent each liquid sample with a hierarchical canonical tuple
\begin{equation}
\mathbf{L} = (L_0, L_1, L_2, L_3),
\end{equation}
where each $L_i$ is a categorical variable capturing a different level of compositional structure:
\begin{itemize}
    \item $L_0$: The main solvent responsible for setting the dielectric relaxation characteristics (such as water-solvent or alcohol-solvent systems).

    \item $L_1$: Main contributor to the mesoscale heterogeneity within the liquid (such as emulsions, suspended phases, and gases within the liquid, e.g., milk fat/protein emulsions, carbonated gas bubbles, or particulate solids).

    \item $L_2$: The major ionic or low electrolyte substance responsible for electrical conductivity losses and relaxation (such as coffee acids/metal ions, phosphoric acid, citric acid, or salts).

    \item $L_3$: Non-ionic substances that affect solvent dynamics and relaxation (such as sugars, syrups, or organics).
\end{itemize}

Each of these liquid mixtures is described using a set of canonical dielectric drivers that follow a hierarchical organization. The word \textit{ingredient} thus refers to dielectric interaction class and not some physical ingredient or macronutrient. These labels can be determined by examining controlled preparation metadata with deterministic rules (for example, $L_2 = \text{Salt}$ in case of a salt-water mixture, or $L_3 = \text{Sugar}$ for a sugar-water mixture).
In case of multiple candidates at a given level, the ingredient with the largest
expected dielectric contribution at the operating frequency is chosen as the dominant
driver. For non-ionic levels ($L_0$, $L_3$) this reduces to mass concentration; for
the ionic level ($L_2$), dominance is assigned to the species with the largest
conductivity contribution at 2.4~GHz rather than by molarity (e.g., phosphoric acid
in cola rather than trace minerals). In all beverages in the current dataset this
assignment is unambiguous because each formulation contains a single clearly dominant
ionic class. The ingredient labels should not be seen as physical measurements and should not be used in any way while making predictions.
The purpose of the labels is that they are used as a supervised signal for the latent electromagnetic interactions driving the radio-frequency observations (the precise train/test data flow is given below).

A canonical classifier predicts each level from the shared RF feature $\mathbf{h}$. We first compute a shared canonical latent:
\begin{equation}
\mathbf{g} = f_{\text{canon}}(\mathbf{h}) \in \mathbb{R}^{B\times d_c},
\end{equation}
where $f_{\text{canon}}(\cdot)$ is a small MLP and $d_c$ is the canonical hidden dimension. We then apply level-specific linear heads:
\begin{equation}
\boldsymbol{\pi}_i = \mathrm{softmax}\!\left(\mathbf{W}_i \mathbf{g} + \mathbf{b}_i\right)
\in [0,1]^{B\times n_i}, \qquad i\in\{0,1,2,3\},
\end{equation}
where $n_i$ denotes the number of classes at level $i$; the model predicts the entire canonical mapping $\hat{\mathbf{L}}$ in a single forward pass from OFDM measurements alone. Separate heads per level provide richer supervision, structured interpretability, and better sample efficiency than a single flat label.

We compute \emph{soft} canonical embeddings from the predicted distributions. For each level $i$, let $\mathbf{E}_i \in \mathbb{R}^{n_i \times d_{\text{canon}}}$ be a learnable embedding table. We compute the expected embedding:
\begin{equation}
\mathbf{e}_i = \boldsymbol{\pi}_i \mathbf{E}_i \in \mathbb{R}^{B\times d_{\text{canon}}}.
\end{equation}
Concatenating all levels yields the canonical conditioning vector:
\begin{equation}
\mathbf{e}_{\text{canon}} = \mathrm{Concat}(\mathbf{e}_0,\mathbf{e}_1,\mathbf{e}_2,\mathbf{e}_3)
\in \mathbb{R}^{B\times 4d_{\text{canon}}}.
\end{equation}

This soft formulation preserves end-to-end differentiability, enabling
gradients from nutrient regression to shape canonical
predictions~\cite{Bengio2013}.

During training, ground-truth canonical labels $\mathbf{L}$ are used \emph{only} as supervision. We apply categorical cross-entropy at each level:
\begin{equation}
\mathcal{L}_{\text{canon}} = \sum_{i=0}^{3} \mathcal{L}_{i},
\qquad
\mathcal{L}_{i} = -\frac{1}{B}\sum_{b=1}^{B} \log \boldsymbol{\pi}_{i}^{(b)}\!\left[L_{i}^{(b)}\right].
\end{equation}

During training, the model receives $\mathbf{X}$ as input and uses $\mathbf{L}$ only to compute $\mathcal{L}_{\text{canon}}$; canonical embeddings are computed from $\boldsymbol{\pi}_i$ (predicted). During testing the model receives only $\mathbf{X}$, predicts $\boldsymbol{\pi}_i$ and $\mathbf{e}_{\text{canon}}$, and outputs nutrients; canonical labels are not provided.

\subsection{Composition-Augmented Macronutrient Regression}
\label{sec:nutrient_regression}

ICOT predicts a continuous macronutrient vector $\hat{\mathbf{y}} \in \mathbb{R}^{B\times D_y}$ (e.g., $D_y=3$ for carbohydrates, fat, protein). We fuse RF and canonical features:
\begin{equation}
\tilde{\mathbf{h}} = \mathrm{Concat}(\mathbf{h}, \mathbf{e}_{\text{canon}}) \in \mathbb{R}^{B\times (d + 4d_{\text{canon}})}.
\end{equation}
A regression head produces the final estimate:
\begin{equation}
\hat{\mathbf{y}} = f_{\text{reg}}(\tilde{\mathbf{h}}),
\end{equation}
where $f_{\text{reg}}(\cdot)$ is an MLP, trained with a robust regression loss (mean absolute error over macronutrient dimensions, defined in Section~\ref{sec:evaluation}).

\subsection{Joint Learning Objective}
\label{sec:joint_objective}

ICOT is trained end-to-end with a weighted multi-task objective:
\begin{equation}
\mathcal{L}_{\text{total}}
  = \alpha_{\text{reg}}\,\mathcal{L}_{\text{reg}}
  + \gamma_{\text{canon}}\,\mathcal{L}_{\text{canon}},
\end{equation}
where $\alpha_{\text{reg}}$ and $\gamma_{\text{canon}}$ balance
nutrient accuracy and structural supervision via a linear warmup:
$\gamma_{\text{canon}}$ decreases from $1.0$ to $0.5$ while
$\alpha_{\text{reg}}$ increases from $0.5$ to $1.0$ over $T_w=20$~epochs
(held fixed thereafter), so the shared encoder first establishes
dielectrically meaningful canonical representations before shifting
emphasis toward nutrient regression~\cite{Caruana1997,Bengio2009}; for
implementation simplicity, the schedule is applied in discrete 5-epoch
increments (four stages), closely approximating the continuous version.
All models are trained with the Adam optimizer~\cite{kingma2015adam}
(learning rate $10^{-4}$, batch size~32) for a maximum of 70~epochs
under early stopping (patience~$= 20$, criterion: validation total loss).

\section{Experimental Setup}
\label{sec:eval_setup}

\begin{figure}[t]
    \centering
    \includegraphics[width=\columnwidth]{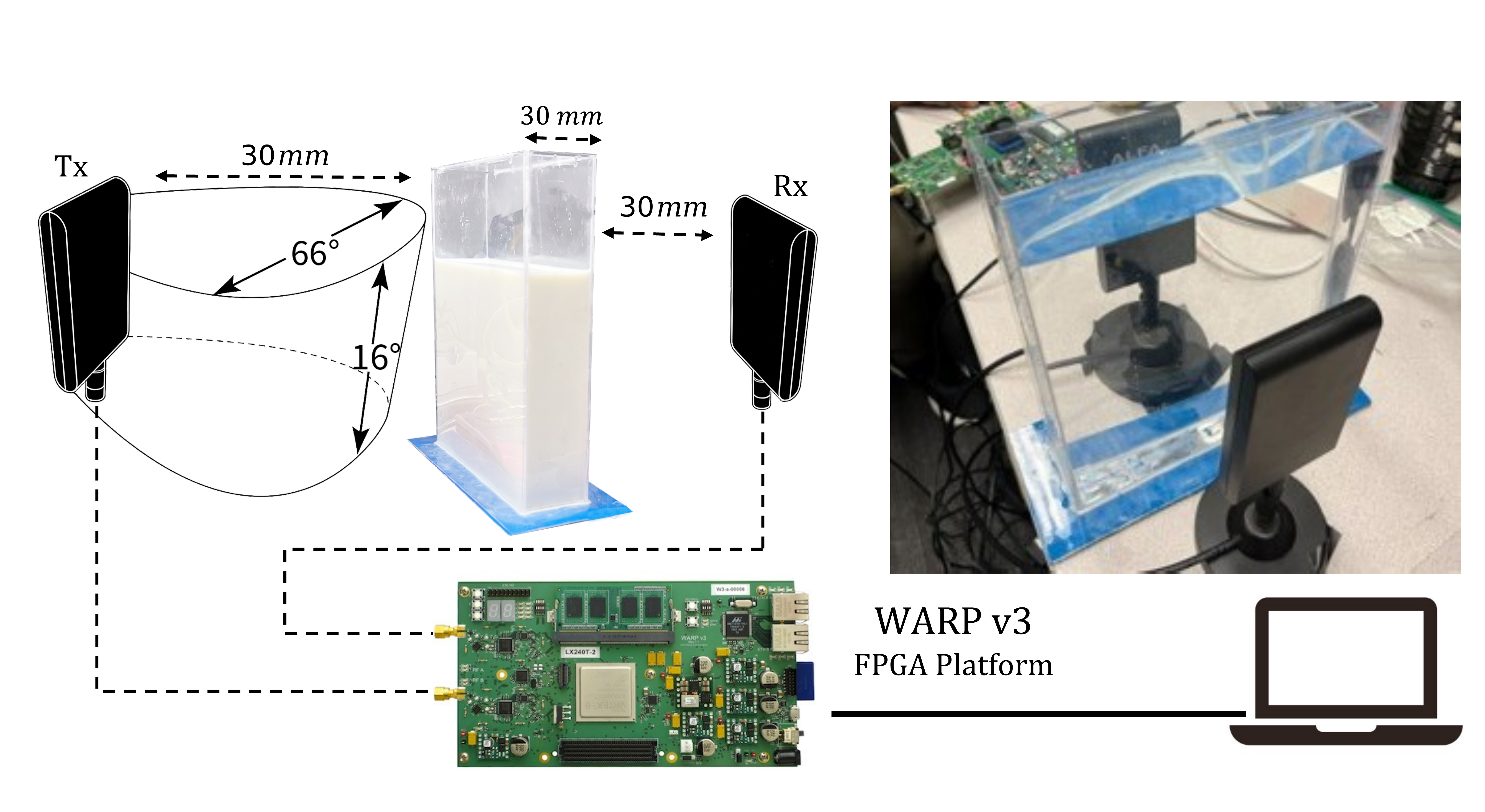}
    \caption{RF liquid-sensing testbed: a WARP v3 SDR drives directional antennas across a fixed-geometry container---annotated schematic (beamwidths marked) alongside a photograph of the physical setup.}
    \label{fig:setup}
\end{figure}

\paragraph{Hardware.}
The testbed (Figure~\ref{fig:setup}) uses a WARP v3 SDR board~\cite{warpProject} (Xilinx Virtex-6 FPGA, two RF interfaces, \SI{40}{MHz} bandwidth, \SI{100}{MSps} ADC) with synchronized transmit and receive chains, distinct from commodity WiFi radios. ALFA Network APA-M25 directional panel antennas (8~dBi gain at 2.4~GHz, 66\textdegree{} horizontal / 16\textdegree{} vertical 3\,dB beamwidth, linear polarization) serve as Tx and Rx, with boresight height aligned to the liquid column's mid-height so that the narrow vertical beamwidth stays fully within the sample across all fill levels tested. A laptop controls the system over Gigabit Ethernet.

\paragraph{RF configuration.}
We transmit OFDM waveforms at \SI{2.4}{GHz} over $C{=}11$ channels. Each 20\,MHz channel uses a 64-bin OFDM grid with 52 active non-null subcarriers; after calibration, we retain the full 64-bin response per channel, producing an $11 \times 64$ channel-response tensor per snapshot~\cite{ofdm_sensing}.

\paragraph{Geometry and containers.}
All liquid samples are kept inside a rectangular glass enclosure that contains a path length of 30\,mm along the direction of the RF beam, has a 2\,mm thick wall on each face, and contains a volume of 2\,L. The Tx and Rx antennas have a separation of 30\,mm from their respective faces, such that the total distance from Tx to Rx is $30 + 2 + 30 + 2 + 30 = 94$\,mm. This isolates channel-response variation to the liquid itself, rather than the path length or container.\paragraph{Calibration.}
Each measurement is normalized by an air reference captured before each session~\cite{calibration_ref}:
\begin{equation}
\label{eq:calibration}
    H_{\mathrm{norm}}(f) = \frac{H_{\mathrm{sample}}(f)}{H_{\mathrm{air}}(f)}\,,
\end{equation}
removing antenna gain, cable loss, and static multipath so that $H_{\mathrm{norm}}(f)$ isolates the liquid's insertion loss and phase shift.

\section{Dataset}
\label{sec:dataset}

\begin{table*}[t]
\centering
\caption{Beverage dataset ($N{=}28{,}420$ captures, 55 classes, plus 1{,}185 ambient-air references), mapped to ABA-BSW/NHANES categories~\cite{aba2021,nhanes2018}. LNCS\,=\,Low/No-Calorie Sweetened; CSD\,=\,Carbonated Soft Drink.}
\label{tab:cse_taxonomy}
\footnotesize
\renewcommand{\arraystretch}{0.8}
\setlength{\tabcolsep}{3pt}
\begin{tabular}{@{} p{0.18\textwidth} p{0.68\textwidth} c r @{}}
\toprule
\textbf{Target Domain} & \textbf{Compositional Variants \& Experimental Matrix} & \textbf{Cls.} & \textbf{$n$} \\
\midrule
Water & Water baseline with ionic (NaCl) and sucrose perturbations, including sugar-in-water variants & 11 & 6{,}040 \\
Regular CSD & Sweetened carbonated beverages spanning cola, lemon--lime, orange, and ginger profiles & 4 & 3{,}110 \\
LNCS CSD & Diet/zero-calorie carbonated beverages spanning cola, lemon--lime, and ginger profiles & 3 & 2{,}450 \\
Coffee-Based Beverages & Brewed coffee baseline; two-level dairy perturbation and two-level sucrose perturbation on the dairy variant & 5 & 810 \\
Tea-Based Beverages & Brewed tea baseline; dairy perturbation and a sucrose perturbation on the dairy variant & 3 & 420 \\
Plain Milk & $3{\times}4$ factorial design with three fat regimes (1\%, 2\%, whole) and four sucrose levels & 12 & 8{,}820 \\
Plant-Based Drinks & Almond milk baseline; two-level sucrose perturbation & 3 & 990 \\
Nutritional Beverages & Two-level protein-fortified formulations & 2 & 340 \\
Energy Drinks & Two base formulations with two-level sucrose perturbation ($2{\times}2$ design) & 4 & 820 \\
Sports Drinks & Two electrolyte base formulations with two-level sucrose perturbation ($2{\times}2$ design) & 4 & 1{,}250 \\
Juices \& Flavored Drinks & Four distinct non-carbonated fruit beverage formulations & 4 & 3{,}370 \\
\midrule
System Reference & Ambient air (empty-container reference capture) & 1 & 1{,}185 \\
\bottomrule
\end{tabular}
\end{table*}

The dataset (Table~\ref{tab:cse_taxonomy}) spans 11 beverage groups plus the system-reference family, each populated with well-controlled compositional variants using structured factorial or staged-perturbation designs (e.g., milk's $3\times4$ fat/sucrose matrix) alongside commercially available formulations. Each capture is a tensor $\mathbf{X}\!\in\!\mathbb{R}^{11 \times 64}$ corresponding to the OFDM channel response (11 channels in the 2.4\,GHz ISM band, 64 subcarriers per channel).
Each capture is annotated with ground-truth carbohydrate, protein, and fat values, acquired directly from manufacturer nutrition labels.%
\footnote{FDA regulations permit a $\pm20\%$ deviation between stated and actual macronutrient values on packaged labels~\cite{jumpertz2013}; for lab-prepared compositional variants (e.g., sucrose additions to a base beverage), ground-truth values were instead computed analytically from ingredient masses and total volumes using established nutrient densities, bypassing label uncertainty.}
As labels often report macronutrient content per serving, we normalized every measure to \textit{grams per 100\,ml}.

\section{Evaluation}
\label{sec:evaluation}

Two complementary evaluation schemes depend on whether the \emph{item} corresponding to the underlying beverage was available during training.
\paragraph{In-Distribution (ID)}The ID scheme splits measurements into training/testing (80/20), ensuring every beverage item appears in both, so testing includes only novel measurements of beverages seen during training.
\paragraph{Out-of-Distribution (OOD)}Generalization to unseen beverage classes is tested via Leave-One-Class-Out (LOCO) validation: for each of the 55 folds, all measurements of one beverage class are held out for testing while the model trains on the remaining classes.
For macronutrient estimation, we report mean absolute error (MAE):
\[
\mathrm{MAE} = \frac{1}{N D_y}\sum_{i=1}^{N}\sum_{j=1}^{D_y}\lvert y_{i,j} - \hat{y}_{i,j}\rvert,
\]
where $D_y{=}3$ is the number of macronutrients; we also report standard deviation across test samples or held-out items.
Per-nutrient MAE fixes $j$ and averages over samples; overall MAE averages over both.

\subsection{Primary Results: Seen-Class (ID) Generalization}
\label{sec:eval_main_results}

\begin{figure*}[t]
    \centering
    \begin{minipage}[t]{\columnwidth}\vspace{0pt}\centering
        \includegraphics[width=\columnwidth]{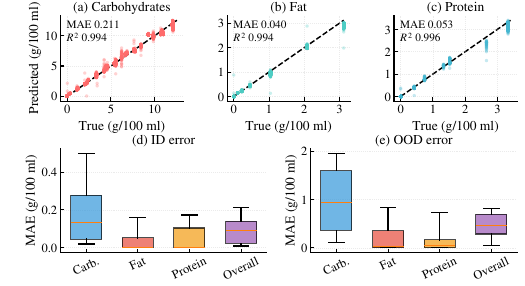}
    \end{minipage}\hfill
    \begin{minipage}[t]{\columnwidth}\vspace{0pt}\centering
        \includegraphics[width=\columnwidth]{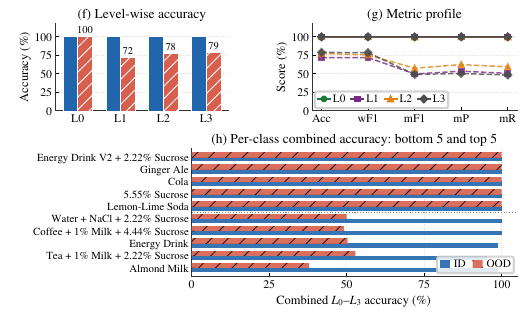}
    \end{minipage}
    \caption{Macronutrient performance and ingredient-level representation. (a)--(c)~ID regression vs.\ ground truth (dashed~=~perfect prediction). (d)--(e)~Per-class error, ID vs.\ OOD (boxes: IQR; whiskers: 10th--90th pct.). (f)~Ingredient accuracy per level. (g)~Precision/recall/F1 per level (solid~=~ID, dashed~=~OOD). (h)~Combined $L_0$--$L_3$ accuracy, five best/worst classes (solid/hatched~=~ID/OOD in f--h).}
    \label{fig:id_performance}
\end{figure*}

Under the ID setting (Figure~\ref{fig:id_performance}), per-nutrient MAE is $0.211$ (carbohydrates), $0.040$ (fats), and $0.053$ (proteins) g/100ml. Fat and protein show low dispersion ($90$th-percentile residuals $0.16$ and $0.17$~g/100ml), while carbohydrate errors are more variable (median $0.13$, whiskers to $0.50$~g/100ml)---the dominant source of residual variance even when the beverage item is seen during training. Figure~\ref{fig:id_performance}(d,e) shows this ID distribution remains tight compared to OOD ($90$th-percentile absolute error $0.21$~g/100ml per sample).

From the representation side, Figure~\ref{fig:id_performance}(f--h) shows ID ingredient classification reaching near-$100\%$ accuracy at every level ($L_0{=}100\%$, $L_1$--$L_3{=}99.9\%$) with correspondingly high precision/recall/F1. Even the five lowest-accuracy beverage categories stay above $80\%$---well-separated, semantically consistent predictions under ID.

\subsection{Primary Results: Unseen-Class (OOD) Generalization}
\label{sec:eval_ood_results}

\begin{table}[t]
\centering
\setlength{\tabcolsep}{4pt}
\renewcommand{\arraystretch}{1.05}
\setlength{\aboverulesep}{1pt}
\setlength{\belowrulesep}{1pt}
\caption{Per-model accuracy and inference time (variants in Section~\ref{sec:ablation}).}
\label{tab:combined_results}
\resizebox{\columnwidth}{!}{
\begin{tabular}{l cccc | cc}
\toprule
\multirow{2}{*}{Model}
  & \multicolumn{2}{c}{ID}
  & \multicolumn{2}{c|}{OOD}
  & \multicolumn{2}{c}{Inference} \\
\cmidrule(lr){2-3}\cmidrule(lr){4-5}\cmidrule(lr){6-7}
 & MAE & RMSE & MAE & RMSE & ms/samp. & samp./s \\
\midrule
ICOT (Ours)         & 0.101 & \textbf{0.200} & \textbf{0.485} & \textbf{0.913} & $0.881{\pm}0.016$ & 1135 \\
ICOT (FiLM)         & \textbf{0.093} & 0.225 & 0.932 & 1.486 & $0.881{\pm}0.016$ & 1135 \\
Vanilla Transf.     & 0.123 & 0.272 & 0.955 & 1.401 & $0.859{\pm}0.019$ & 1164 \\
Transf. (No Ingr.)  & 0.169 & 0.314 & 1.064 & 1.677 & $0.861{\pm}0.016$ & 1161 \\
CNN Baseline        & 0.399 & 0.752 & 1.121 & 1.454 & $0.513{\pm}0.016$ & 1951 \\
\bottomrule
\end{tabular}}
\end{table}

Table~\ref{tab:combined_results} and Figure~\ref{fig:id_performance}(d,e) show all models' MAE/RMSE increase substantially under LOCO relative to ID (for ICOT, both roughly quadruple), unevenly across categories. OOD distributions are wider and more positively skewed, with carbohydrate errors to $2.0$~g/100ml at the whisker tips (worst-class MAE $2.95$)---concentrated on a smaller number of unseen beverages---while fat/protein stay fairly accurate (median $<0.05$) and carbohydrates remain most volatile (median $\approx 0.94$).

The long-tail is concentrated in a small number of classes---2\% Milk+6.66\% Sucrose, 1\% Fat Milk, and Tea+1\% Milk+2.22\% Sucrose---each with the highest carbohydrate residual (MAE $2.95$, $2.74$, $2.16$~g/100ml). 1\% Fat Milk is the hardest class overall (MAE $2.13$): as the only unsweetened member of its fat regime, it is misread as non-dairy once held out, misplacing both lipid and protein (MAE $0.89$, $2.77$). High-sucrose dairy/tea mixtures see elevated error because added sucrose shifts their dielectric response into a region overlapping seen formulations, leaving carbohydrate under-determined; diet/zero-calorie beverages, with water-like responses and high training representation, are easiest. Class-weighted mean OOD MAE is $0.510$~g/100ml, slightly above the per-sample $0.485$ in Table~\ref{tab:combined_results} since the largest-error classes (dairy) contribute fewer samples than well-predicted carbonated/juice classes.

Under LOCO (Figure~\ref{fig:id_performance}(f--h)), ingredient classification accuracy at $L_1$--$L_3$ drops from $99.9\%$ (ID) to $71.8\%$, $77.5\%$, and $79.0\%$ respectively, while $L_0$ stays at $100\%$---the structural and ionic/non-ionic levels most directly linked to macronutrient content are hardest to infer for unseen formulations.

\subsection{Ablation Study: Architectural Validation}
\label{sec:ablation}

We perform an ablation study evaluating (i) transformer architecture, (ii) ingredient-level supervision, and (iii) conditioning technique separately, using the same training protocol as the full model (Section~\ref{sec:joint_objective}).
\paragraph{Compared Model Variants}
\label{sec:ablation:variants}
We evaluate five architectures:
\begin{itemize}
    \item \textit{ICOT} (ours): Transformer with ingredient-level 
    supervision and concatenation-based fusion at the regression head.
    \item \textit{ICOT (FiLM)}: Same backbone as ICOT, but with 
    FiLM-based~\cite{perez2018film} feature modulation.
    \item \textit{Vanilla Transformer}: Transformer regressor on 
    RF features only, without any ingredient branch.
    \item \textit{Transformer (No Ingr.)}: ICOT architecture with 
    the ingredient path disabled, capacity-matched to ICOT.
    \item \textit{CNN Baseline}: 1D CNN regressor without a 
    Transformer or ingredient branch.
\end{itemize}

ICOT (FiLM) modulates the Transformer feature via an affine transform
$g_\gamma(\hat{z}) \odot \mathbf{h} + g_\beta(\hat{z})$ driven by a 32-dim
ingredient bottleneck $\hat{z}$~\cite{perez2018film}, risking attenuation
through multiplicative gating. ICOT instead concatenates $\hat{z}$ onto
$\mathbf{h}$, preserving all RF features.
Both ablation variants share this 32-dim bottleneck,
whereas the full ICOT model uses the richer 4-level canonical embedding of
Section~\ref{sec:canonical_mapping} ($d + 4d_{\text{canon}}$). The ICOT (ours) vs.\
ICOT (FiLM) comparison in Table~\ref{tab:combined_results} therefore conflates
supervision richness with conditioning mechanism, whereas the two ablation
variants isolate the mechanism cleanly.

Table~\ref{tab:combined_results} yields three observations: (1)~the CNN baseline
has substantially higher error than all Transformer variants, confirming
convolutional local features cannot capture the non-local spectral
dependencies of dielectric interactions; (2)~variants without ingredient
conditioning (Vanilla Transformer, Transformer (No Ingr.)) underperform ICOT
in both ID and OOD, mostly on carbohydrate estimation; and (3)~ICOT achieves
the strongest OOD accuracy and lowest ID RMSE, tied with ICOT (FiLM) on ID
MAE while reducing OOD MAE by $\sim\!48\%$. FiLM's substantially higher OOD
RMSE suggests multiplicative gating amplifies sensitivity to distributional
shift~\cite{geirhos2020shortcut}, a known failure mode of auxiliary-conditioned
models.

\subsection{Ablation Study: Ingredient Conditioning Evaluation}
\label{sec:conditioning_eval}

Ingredient conditioning aims to disambiguate RF responses, but
auxiliary signals can act as fragile
shortcuts---yielding strong in-distribution gains while failing
under distribution shift~\cite{geirhos2020shortcut}. We evaluate it along two axes: (i) functional dependence via embedding-shuffle
analysis, and (ii) functional reliance via gradient attribution. The shuffle
breaks the statistical association between embeddings and RF inputs but is
not a do-calculus intervention, so results establish \emph{functional}
dependence rather than strict causal necessity.

\paragraph{Embedding-Shuffle Analysis.}
We shuffle predicted ingredient embeddings at inference time while
keeping RF inputs fixed, breaking ingredient--signal correspondence
without altering marginal distributions. As shown in 
Table~\ref{tab:shuffle_combined}, this causes a large ID 
degradation: MAE increases by $+660\%$ for ICOT and $+1591\%$ for
ICOT (FiLM). Under OOD-LOCO, however, shuffling produces 
negligible degradation ($\approx 0\%$), indicating that ingredient 
conditioning does not generalize to unseen compositions.

\begin{table}[t]
\centering
\setlength{\tabcolsep}{4pt}
\renewcommand{\arraystretch}{1.05}
\setlength{\aboverulesep}{1pt}
\setlength{\belowrulesep}{1pt}
\caption{Embedding-shuffle and gradient-attribution analysis (ID, OOD-LOCO).}
\label{tab:shuffle_combined}
\resizebox{\linewidth}{!}{
\begin{tabular}{l cccc | cccccccc}
\toprule
\multirow{3}{*}{Model}
  & \multicolumn{4}{c|}{Embedding Shuffle}
  & \multicolumn{8}{c}{Gradient Attribution} \\
\cmidrule(lr){2-5}\cmidrule(lr){6-13}
  & \multicolumn{2}{c}{ID} & \multicolumn{2}{c|}{OOD}
  & \multicolumn{4}{c}{ID} & \multicolumn{4}{c}{OOD} \\
\cmidrule(lr){2-3}\cmidrule(lr){4-5}\cmidrule(lr){6-9}\cmidrule(lr){10-13}
  & Orig. & Shuf. & Orig. & Shuf.
  & $\rho$ & Std & $\|\nabla_{\hat{z}}\|$ & $\|\nabla_{x}\|$
  & $\rho$ & Std & $\|\nabla_{\hat{z}}\|$ & $\|\nabla_{x}\|$ \\
\midrule
ICOT (Ours) & 0.101 & 0.769\,{\scriptsize(+660\%)}  & 0.485 & 0.482\,{\scriptsize($\approx$0\%)}
            & 0.194 & 0.141 & 0.500 & 4.687
            & 0.092 & 0.063 & 0.649 & 11.359 \\
ICOT (FiLM) & 0.093 & 1.578\,{\scriptsize(+1591\%)} & 0.932 & 0.931\,{\scriptsize(-0.16\%)}
            & 0.405 & 0.517 & 0.888 & 5.530
            & 0.174 & 0.139 & 1.425 & 19.196 \\
\bottomrule
\end{tabular}}
\end{table}

\paragraph{Gradient Attribution.}
We quantify prediction dependence on ingredient embeddings via a 
gradient sensitivity ratio:
\[
\rho = \frac{\|\nabla_{\hat{z}}\hat{y}\|}{\|\nabla_{x}\hat{y}\|},
\]
where $x$ denotes RF features and $\hat{z}$ the predicted
ingredient embedding. FiLM-based conditioning induces $2.09\times$ higher ID
ingredient sensitivity than concatenation (Table~\ref{tab:shuffle_combined}),
yet gains no accuracy advantage---marginally ahead on ID MAE ($0.093$ vs.\
$0.101$) but worse on ID RMSE and both OOD metrics, suggesting excessive
reliance is detrimental. Sensitivity drops sharply for both models under
OOD-LOCO, reinforcing that conditioning is not functionally relied upon when
generalization is required---the same asymmetry as the embedding-shuffle
results.
\subsection{Practical and Compositional Variability}
\label{sec:eval_robustness_mixtures}

We evaluate ICOT's robustness under practical deployment conditions, 
probing sensitivity to thermal variability, mixture complexity, and 
polar additive combinations.

\paragraph{Thermal Conditions.}

\begin{figure}[t]
    \centering
    \includegraphics[width=\columnwidth]{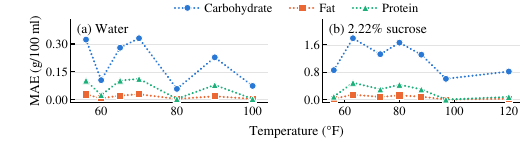}
    \caption{MAE across temperatures for (a) water and (b) 2.22\% (w/v) sucrose.}
    \label{fig:temp_mae}
\end{figure}

To assess thermal robustness, we acquired RF spectra from water and 2.22\% (w/v) sucrose solutions across controlled temperatures, held out from training and evaluated under OOD-LOCO (Fig.~\ref{fig:temp_mae}). Water-based MAE stays low across all temperatures (carbohydrate $\leq 0.35$; fat/protein $<0.15$~g/100ml), but sucrose-solution MAE rises substantially near 62\textdegree F and 80\textdegree F ($\approx 1.78$ and $1.65$~g/100ml) before partially recovering. Fat stays consistent throughout ($<0.15$) while protein peaks near 62\textdegree F ($\approx 0.50$)---establishing thermal variation as a structured, carbohydrate-concentrated source of distributional shift.

\paragraph{Mixed-Liquid Evaluation.}

\begin{figure*}[t]
    \centering
    \includegraphics[width=\textwidth]{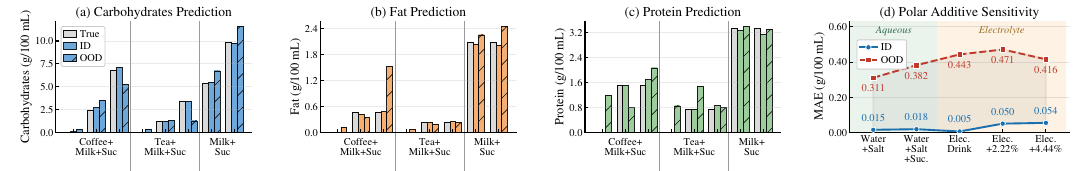}
    \caption{Macronutrient prediction under ID and OOD-LOCO. (a)--(c)~Mixture-complexity progressions (coffee, tea, milk); labels name the final mixture, steps run left to right in increasing complexity. (d)~Error for polar additive combinations (aqueous, then electrolyte).}
    \label{fig:progression_comparisons}
\end{figure*}

This progression design probes whether ICOT tracks compositional change rather than memorized identity: each step adds one precisely known ingredient to the same base liquid, so an identity-lookup model would not shift its prediction within a family, whereas a composition-tracking model should.
As Figure~\ref{fig:progression_comparisons} shows, ID predictions do shift with each addition, tracking the known composition change, and error grows with mixture complexity. Under OOD, the zero-nutrient progressions (black coffee/tea) exhibit a spurious protein reading ($\approx 1.2$, $0.9$~g/100ml against a ground truth of zero), showing the model reverts to the training-set mean once the base liquid is unseen. Sucrose additions produce systematic carbohydrate underestimation (e.g., $\approx$5 vs.\ $\approx$7~g/100ml true for coffee+milk+sucrose), and milk progressions show the largest OOD fat error ($\approx 2.4$ vs.\ $\approx 2.1$~g true) while protein stays close to ground truth. These results indicate the model learns stable single-ingredient representations, with unseen multi-component combinations remaining the primary avenue for improved generalization.

\paragraph{Sensitivity to Polar Additives.}
As shown in Figure~\ref{fig:progression_comparisons}(d), ID errors remain uniformly low across all compositions (MAE $\leq 0.054$~g/100ml), while OOD errors are substantially higher and composition-dependent. Within the aqueous group, adding sucrose increases OOD error (0.311$\to$0.382~g/100ml), indicating dissolved sucrose rather than ionic content drives the shift. The electrolyte group shows a non-monotonic trend, suggesting a non-linear electrolyte--sucrose interaction.

\paragraph{Inference Time.}

Inference latency, averaged over 1{,}000 forward passes on the ID test set (CPU only), is $0.859$--$0.881$~ms/sample for all Transformer variants---confirming minimal overhead from ingredient conditioning---versus a faster but substantially less accurate CNN baseline ($0.513$~ms).

\subsection{Practical Error Interpretation}
\label{sec:error_interp}

Carbohydrate counting commonly uses a 15~g serving exchange as a reference unit~\cite{roversi2022}; the FDA permits $\pm 20\%$ deviation on packaged labels, and \citet{jumpertz2013} measured actual snack-label deviations averaging 4\%--8\% per item.

Against these benchmarks, ICOT's ID MAE of $0.101$~g/100ml represents less than $0.7\%$ of the 15~g exchange---clinically negligible.
The \emph{average} OOD MAE ($0.485$~g/100ml, $\approx 3.2\%$ of the same threshold) is comparable to accepted label deviations, but conceals a long tail: compositionally singular items---1\% Fat Milk (MAE $2.74$) and high-sucrose dairy mixtures (MAE $2.95$~g/100ml)---reach clinically significant errors, so closing these specific compositional gaps is a prerequisite for OOD deployment. The average-MAE comparison applies to in-distribution liquids only.
These OOD errors are not arbitrary---they are identifiable gap classes arising from compositional singularity and spectral confusability (Section~\ref{sec:conclusion} elaborates both error sources and the resulting research agenda). Outside these gap classes, ICOT's performance already matches FDA-permitted label tolerances.

\section{Related Work}
\label{sec:related_work}

Wearable methods deduce nutrition from behavior-based information: acoustic sensors near the mouth/throat recognize chewing and swallowing~\cite{chewpin2021}, identifying \emph{eating occurrences} rather than \emph{nutritional value}, so compositionally distinct liquids (e.g., water vs.\ protein beverages) remain indistinguishable.
Optical alternatives---smartphone-LiDAR viscosity~\cite{chan2022lidarliquid}, MobiLyzer~\cite{mobilyzer2025}, hyperspectral inference~\cite{xu2016development}---require visual contact with the liquid surface and bespoke hardware, becoming unreliable for opaque containers or unlabeled home-prepared beverages.

Liquid classification from radio signals has emerged as a distinct sensing modality within the broader landscape of RF-based material sensing~\cite{chen2024wireless}, spanning UWB permittivity sensing (\textit{LiquID}~\cite{10.1145/3210240.3210345}), commodity WiFi CSI (\textit{WiMi}~\cite{feng2018material}, \textit{WiEps}~\cite{song2020wieps}), millimeter-wave classification (\textit{FG-LiquID}~\cite{liang2021fgliquid}, \textit{LiqRay}~\cite{shang2022liqray}, \textit{PackQuID}~\cite{shang2023packquid}, \textit{mmTaster}~\cite{liang2024mmtaster}), and RFID backscatter (\textit{RF-EATS}~\cite{ha2020rfeats}).
Table~\ref{tab:rf_comparison} summarizes each system's modality, reported accuracy metric, and portability alongside optical alternatives (\textit{MobiLyzer}~\cite{mobilyzer2025}) and ICOT. Almost all report identity or property-classification accuracy rather than absolute per-volume macronutrient quantities, and ICOT trades the portability of commodity radios for this quantitative output.

\begin{table}[t]
\centering
\caption{Prior RF- and optical-based liquid-sensing systems.}
\label{tab:rf_comparison}
\footnotesize
\renewcommand{\arraystretch}{0.85}
\setlength{\tabcolsep}{2.5pt}
\resizebox{\columnwidth}{!}{%
\begin{tabular}{@{} l l p{0.42\columnwidth} c @{}}
\toprule
\textbf{System} & \textbf{Modality} & \textbf{Reported Metric} & \textbf{Port.} \\
\midrule
WiNE~\cite{wine2022} & $\mu$wave leakage & MAE $\leq$5\% nutrient; $r{\approx}0.97$ cal. & No \\
RFTemp~\cite{rftemp2021} & $\mu$wave leakage & 5\textdegree C mean temp. error & No \\
LiquID~\cite{10.1145/3210240.3210345} & UWB & 9\% median permittivity err. (33 liquids) & Yes \\
WiMi~\cite{feng2018material} & WiFi CSI & 95--99\% acc. (10 liquids) & Yes \\
WiEps~\cite{song2020wieps} & WiFi/OFDM & 4.0\%/8.9\% permittivity/cond. err. & Yes \\
FG-LiquID~\cite{liang2021fgliquid} & mmWave & 97\% acc. (30 liquids) & Yes \\
LiqRay~\cite{shang2022liqray} & mmWave & 92.9\% acc., 1\% conc. res. & Yes \\
PackQuID~\cite{shang2023packquid} & mmWave & 86--87\% acc. (9 liquids) & Yes \\
RF-EATS~\cite{ha2020rfeats} & RFID backscatter & $>$90\% acc. & Yes \\
mmTaster~\cite{liang2024mmtaster} & mmWave & 90.8\% acc., 0.2\% v/v res. & Yes \\
MobiLyzer~\cite{mobilyzer2025} & Optical & Qualitative (adulteration/origin) & Yes \\
\textbf{ICOT (ours)} & Sync. OFDM & \textbf{0.101 / 0.485} g/100\,ml MAE (ID/OOD, 55 beverages) & No \\
\bottomrule
\end{tabular}}
\end{table}

ICOT implements hard parameter sharing~\cite{Caruana1997}: a Transformer encoder is jointly trained for dielectric driver inference and macronutrient regression. Theory formalizes why auxiliary supervision aids OOD robustness when auxiliary labels inform a shared latent space~\cite{xie2021innout}: auxiliary \emph{output} labels help, whereas auxiliary \emph{input} labels harm it if the input-output relationship changes. We use ingredient-level dielectric driver information as an auxiliary output during training only, motivating concatenation over affine FiLM-style~\cite{perez2018film} conditioning to avoid the shortcut-reliant failure mode of~\cite{geirhos2020shortcut} analyzed in Section~\ref{sec:ablation}.

\section{Conclusion}
\label{sec:conclusion}

This work establishes that active OFDM measurements at 2.4~GHz carry sufficient dielectric information for contact-free, label-free macronutrient estimation in liquid foods---a basis for RF-based dietary sensing as an alternative to label-dependent and vision-based methods. The current synchronized-hardware, fixed-volume testbed, however, is a feasibility study rather than a deployable commodity sensor.
The physical sensing chain has two stages---RF signals couple with dielectric constituents, governed by ingredient composition---and ICOT aligns its inference architecture with this chain, decomposing composition identification from composition-conditioned regression. This alignment is empirically active rather than incidental.

ICOT achieves high ID ingredient classification accuracy ($L_0$--$L_3 \geq 99.9\%$) and overall MAE of $0.101$~g/100ml, versus $0.485$~g/100ml under OOD-LOCO---outperforming all baselines including FiLM-conditioned and ingredient-free variants. Embedding-shuffle intervention confirms the canonical pathway is functionally active (ID error $+660\%$ when shuffled) rather than circumvented, and that concatenation-based fusion is more robust than FiLM under distribution shift.

OOD errors fall into two categories: compositional uniqueness (no training analogue, e.g., 1\% Fat Milk, the only unsweetened member of its fat regime) and spectral ambiguity (dissolved sugar overlapping seen formulations' dielectric spectra, e.g., high-sucrose milk/tea mixtures). These set a clear research agenda: broader carbohydrate concentration coverage, closer-to-isodielectric training beverages, and portable non-synchronized hardware. High-concentration carbohydrate generalization remains the key outstanding problem for the diabetes use case motivating this work.

\bibliographystyle{IEEEtran}
\bibliography{references}

\end{document}